\documentclass[3p,times,procedia]{elsarticle}
\usepackage{nupha_ecrc}

\volume{00}

\firstpage{1}

\journalname{Nuclear Physics A}

\runauth{}

\jid{nupha}

\jnltitlelogo{Nuclear Physics A}

\usepackage{amssymb}

\usepackage{lineno}

\usepackage[figuresright]{rotating}
\usepackage{wrapfig}

\begin{document}

\begin{frontmatter}



\dochead{}

\title{Charge-dependent anisotropic flow in Cu+Au collisions}


\author{Takafumi Niida for the STAR Collaboration\footnotemark[1]}

\address{Wayne State University, 666 W. Hancock, Detroit, MI 48201}

\begin{abstract}
We present the first measurements of charge-dependent directed flow in Cu+Au collisions at $\sqrt{s_{_{NN}}}$= 200 GeV.
The directed flow has been measured as functions of the transverse momentum and pseudorapidity with the STAR detector. The results show a small but finite difference between positively and negatively charged particles. The difference is qualitatively explained by the patron-hadron-string-dynamics (PHSD) model including the effect of the electric field, but much smaller than the model calculation, which indicates only a small fraction of all final quarks are created within the lifetime of the initial electric field.
Higher-order azimuthal anisotropic flow is also presented up to the fourth-order for unidentified charged particles and up to the third-order for identified charged particles ($\pi$, $K$, and $p$). For unidentified particles, the results are reasonably described by the event-by-event viscous hydrodynamic model with $\eta/s$=0.08-0.16. The trends observed for identified particles in Cu+Au collisions are similar to those observed in symmetric (Au+Au) collisions.
\end{abstract}


\end{frontmatter}


\footnotetext[1]{A list of members of the STAR Collaboration and acknowledgements can be found at the end of this issue.}

\section{Introduction}
At the early stage of a non-central heavy-ion collision, a strong
magnetic field perpendicular to the reaction plane is created. In
asymmetric Cu+Au collisions, due to the difference in the number of
spectators, not only the magnetic field but also a strong electric
field would exist pointing along the reaction plane from the
Au-nucleus to Cu-nucleus. The lifetime of the electric field might be
short, of the order of a fraction of a fm/$c$. The quarks and antiquarks
that have been already produced at this time would experience the
Coulomb force, which results in a charge dependence of particle directed
flow~\cite{hirono,voronyuk}.  Thus, the measurement of the
charge-dependent directed flow in Cu+Au collisions provides an opportunity to
test different quark (charge) production scenarios, e.g. two-wave quark
production~\cite{2wave1,2wave2}, and shed light on the (anti-)quark
production mechanism in heavy-ion collisions. Understanding the time
evolution of the quark densities in heavy-ion collisions is also very
important for detailed theoretical predictions of the Chiral Magnetic
Effect and Chiral Magnetic Wave, for which various experiments are
actively searching.
In these proceedings, the charge-dependent directed flow in Cu+Au
collisions at $\sqrt{s_{_{NN}}}$= 200 GeV measured with the STAR
detector is presented. Results of higher-order flow are also
presented.
\section{Analysis method}
Azimuthal anisotropies were measured with the event plane method
defined below:
\begin{equation}
v_{n} = \langle \cos n(\phi-\Psi_{n}) \rangle / {\rm Res}\{\Psi_{n}\},
\end{equation}
where $\phi$ is azimuthal angle of particles and $\langle \rangle$
means average over all particles in the events of the same centrality bins.  The $\Psi_{n}$
denotes n$^{\rm th}$-order event plane.
The first-order event plane was reconstructed with the Zero Degree Calorimeter (ZDC). 
The ZDC measures spectator neutrons and thus would minimize non-flow effects such as those from the
momentum conservation.  For higher harmonics measurements, the event planes were
reconstructed from charged tracks (0.15$<$$p_{T}$$<$2 GeV/$c$) reconstructed
in the Time Projection Chamber (TPC) and the Endcap Electro-Magnetic
Calorimeter (EEMC). In case of using the TPC, charged tracks were divided
into two subevents (-1$<$$\eta$$<$-0.4 and 0.4$<$$\eta$$<$1) and
$v_{n}$ of charged particles of interest was measured with an
$\eta$-gap of 0.4 using the event plane method (e.g. particles of interest are taken from 0$<$$\eta$$<$1 when using the subevent from the backward angle). The results from both subevents are consistent and the average of two measurements was used as final results. 
The event plane resolution Res\{$\Psi_{n}$\} was estimated
by three subevents method~\cite{TwoSub}.
Systematic uncertainties were estimated by varying event z-vertex and
track quality cuts. The effect of the event plane determination was
also taken into account in the systematic uncertainty. For
higher-order $v_{n}$, the scalar product method~\cite{SP} was also
tested just as a cross-check.

\section{Results}
\footnotetext[2]{Note that the figures in these proceedings have been updated since the presentation to
  account for a software issue in the calculation of the event plane resolution. The final results on $v_{1}$ are quantitatively close to those presented at the conference and had no impact on the physics conclusions. The $v_{2}$ and $v_{3}$ in peripheral collisions become larger after this correction.}
\begin{figure}[htb]
\begin{center}
\includegraphics[width=0.95\linewidth,angle=0]{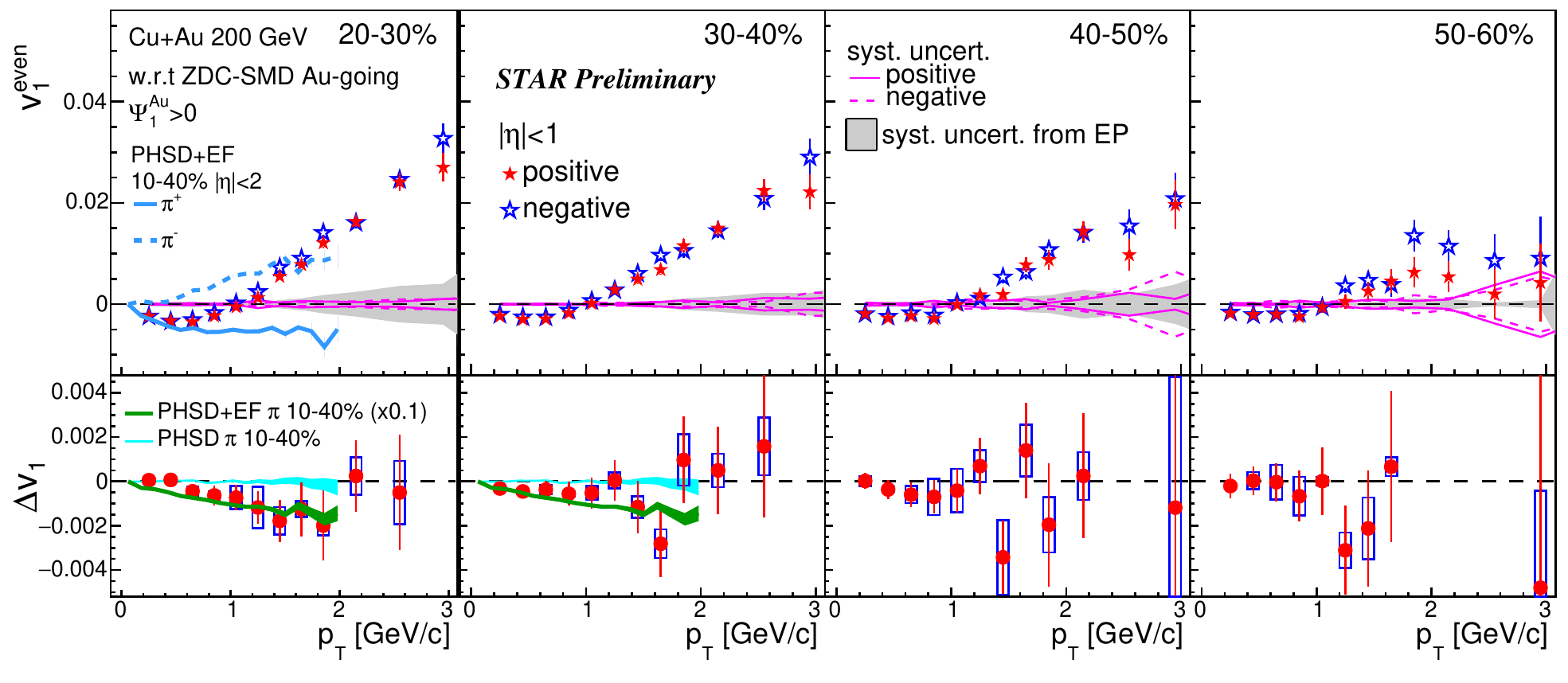}
\caption{$v_{1}^{even}$ of positive and negative particles as
  a function of $p_{T}$ in four centrality bins and the difference
  between both charges, $\Delta v_{1}$. The PHSD model calculations
  with and without the initial electric field (EF)~\cite{voronyuk} are
  compared. The model calculation of $\Delta v_{1}$ with the EF is scaled by 0.1. See the text for the definition of positive direction of $v_{1}$ ($\Psi_{1}$). This plot has been updated since the presentation\protect\footnotemark[2].}
\label{fig:v1pt}
\end{center}
\end{figure}
Figure~\ref{fig:v1pt} shows $v_{1}$ of positive ($h^{+}$) and
negative ($h^{-}$) charged particles as a function of $p_{T}$ in four
centrality bins\footnotemark[2], where $v_{1}$ is measured with
respect to the spectator plane in Au-going side and the sign of the
$\Psi_{1}^{\rm Au}$ is defined to be positive. In asymmetric collisions, the magnitude of $v_{1}$ is no longer symmetric over the pseudorapidity unlike symmetric collisions, therefore the even component of $v_{1}$ is measured in this analysis.
The $v_{1}$ at $p_{T}$$<$1 GeV/$c$ has negative value and positive at the higher $p_{T}$, 
which means more low (high) $p_{T}$ particles are emitted to the direction of Cu (Au) spectator.
Bottom panels of Fig.~\ref{fig:v1pt} show the difference between both charges, $\Delta
v_{1}=v_{1}^{h^{+}}-v_{1}^{h^{-}}$.  In 20-40\% centrality, the
$\Delta v_{1}$ seems to be negative in $p_{T}$$<$2 GeV/$c$, which is
qualitatively consistent with the expectation from the initial
electric field (EF), i.e. more positively charged particles would move
to the direction of the EF and negatively charged particles move to the
opposite side. 

\begin{wrapfigure}[22]{r}[1mm]{0.44\linewidth}
\begin{center}
\includegraphics[width=\linewidth,angle=0]{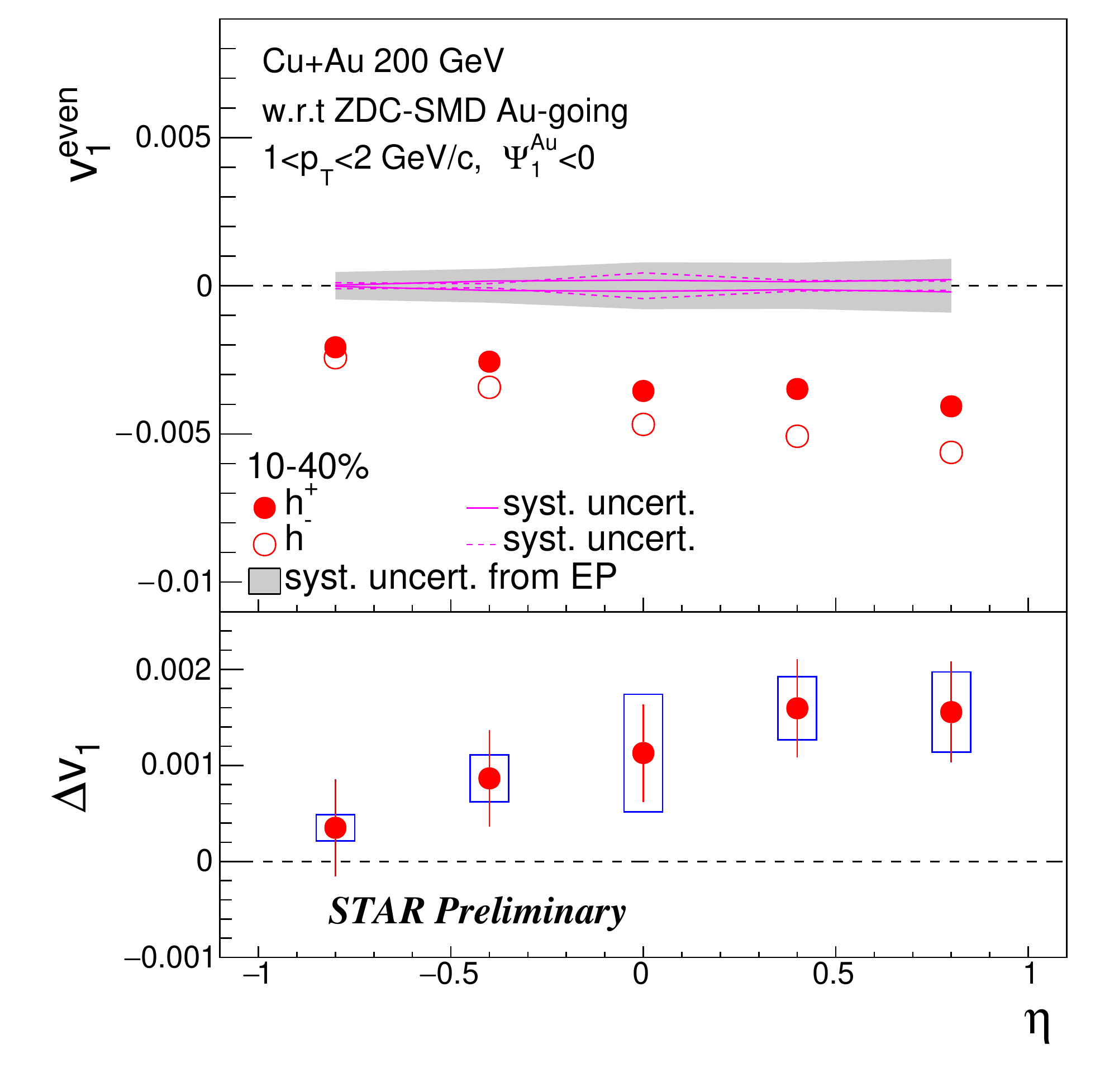}
\caption{$v_{1}$ of positive and negative  particles and
  $\Delta v_{1}$ as a function of $\eta$ in 10-40\% centrality. This plot has been updated since the presentation\protect\footnotemark[2].}
\label{fig:v1eta}
\end{center}
\end{wrapfigure}
The parton-hadron-string-dynamics (PHSD) model calculations with and without the effect of the
EF~\cite{voronyuk} are compared to the data, where the $\Delta v_{1}$ for the calculation including the effect of the EF
is scaled by 0.1. The model assumes that all electric charges are
affected by the EF, resulting in a large separation of $v_{1}$ between
positive and negative charges as shown in the upper left panel of Fig.~\ref{fig:v1pt}.  
The $\Delta v_{1}$ is smaller than the model prediction, which indicates
that the electric charges existing within the life time of the EF
($\sim$0.25 fm/$c$) are much smaller than that of all quarks created in the collisions.

Figure~\ref{fig:v1eta} shows $v_{1}$ and $\Delta v_{1}$ as a function of $\eta$ in 10-40\%
centrality\footnotemark[2], where $p_{T}$ is integrated over
1$<$$p_{T}$$<$2~GeV and the sign of $\Psi_{1}^{\rm Au}$ is defined to be
negative (opposite to Fig.~\ref{fig:v1pt}). The $v_{1}$ charge
separation is clearly seen in $|\eta|$$<$1 and $\Delta v_{1}$ increases with $\eta$,
although the magnitude of $v_{1}$ also changes with $\eta$.  

\begin{figure}[hbt]
\begin{minipage}{0.56\hsize}
\begin{center}
\includegraphics[width=0.98\textwidth,angle=0,trim=0 0 0 0]{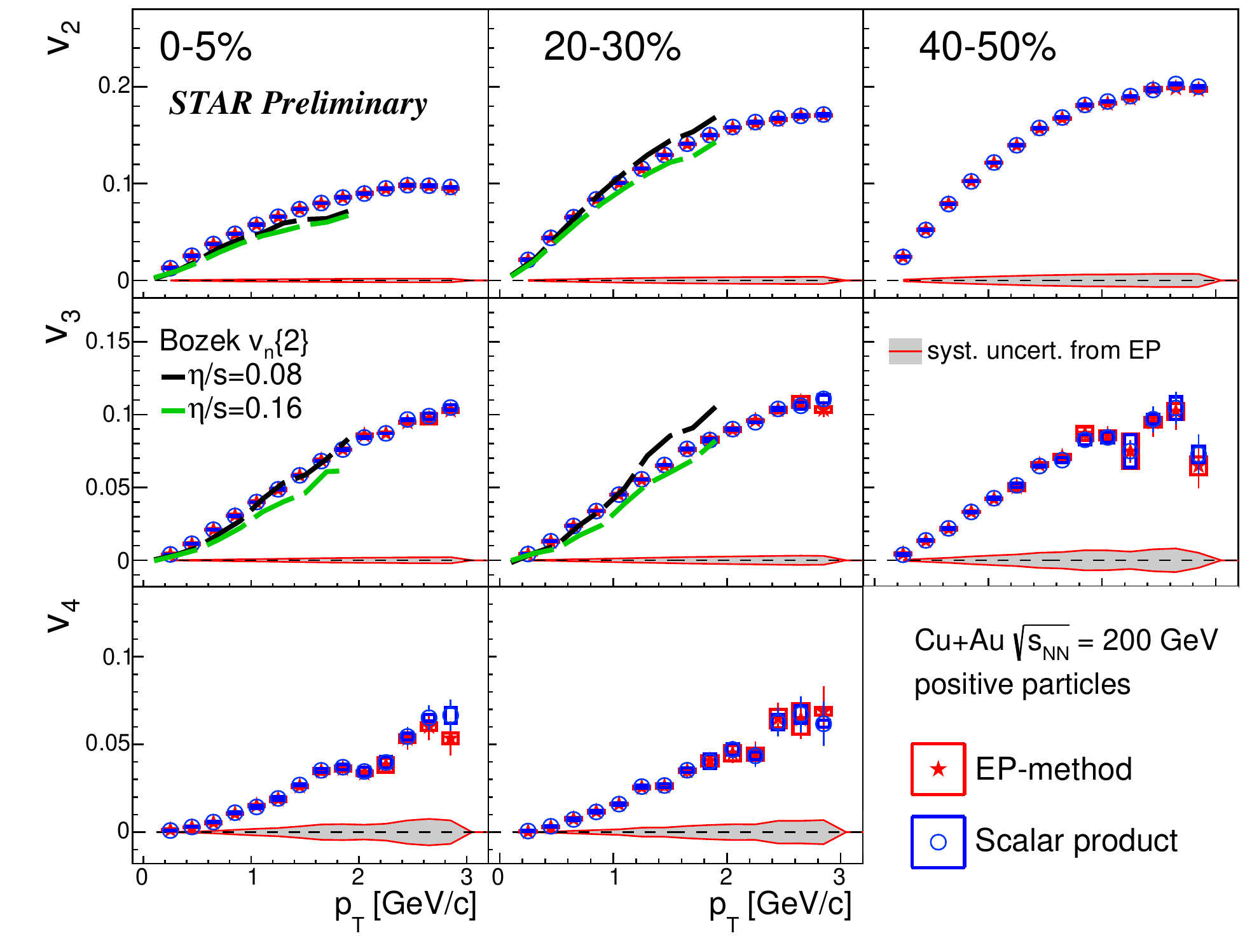}
\caption{$v_{2}$, $v_{3}$, and $v_{4}$ as a function of $p_{T}$ in
  0-5\%, 20-30\%, and 40-50\% centrality bins measured with the event plane method and the scalar product method. 
  Calculations from the event-by-event viscous hydrodynamic model~\cite{bozek} are compared. 
  This plot has been updated since the presentation\protect\footnotemark[2].}
\label{fig:vn}
\end{center}
\end{minipage}
\begin{minipage}{0.43\hsize}
\begin{center}
\includegraphics[width=0.96\textwidth,angle=0,trim=0 0 0 0]{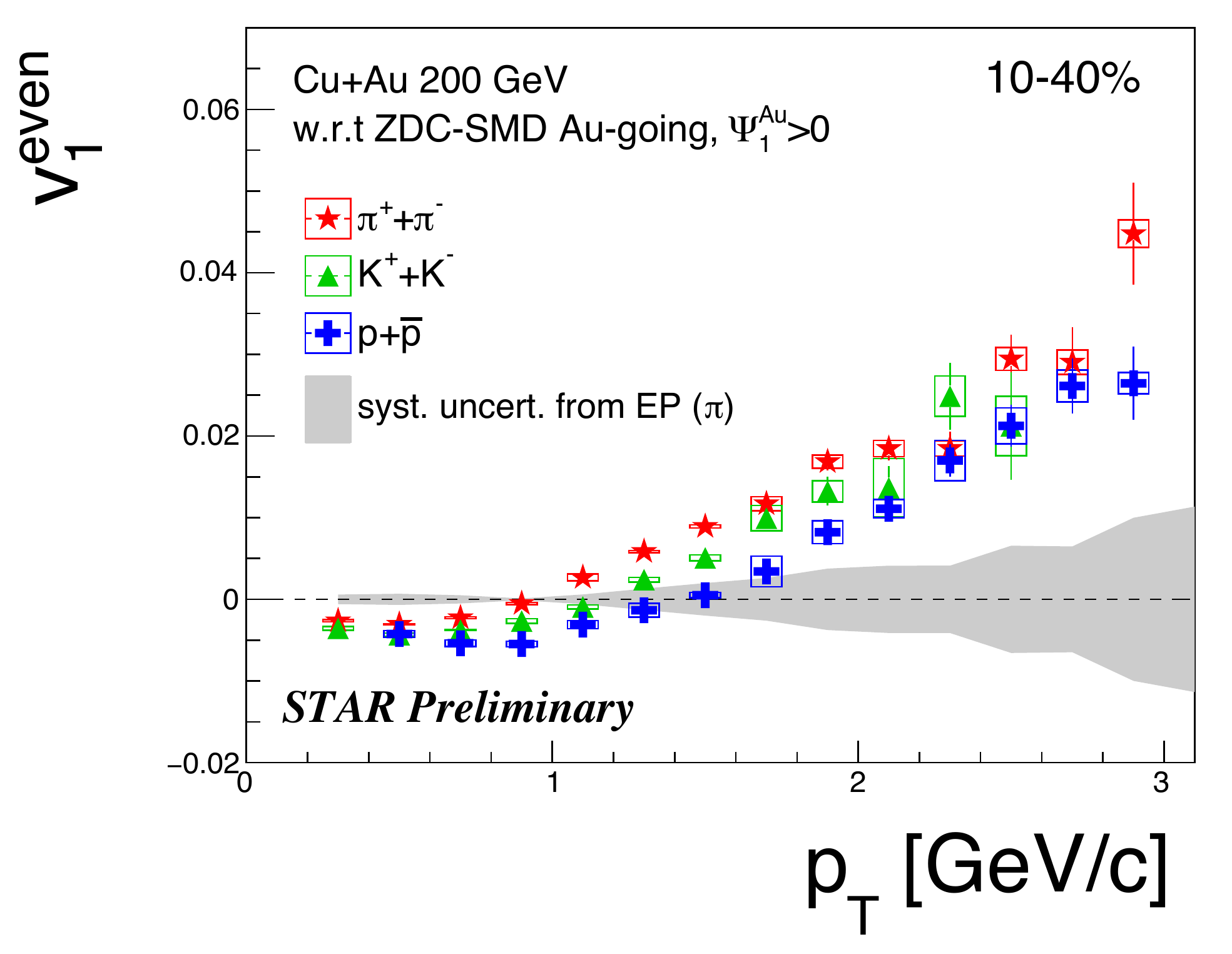}
\caption{$v_{1}$ of $\pi^{\pm}$, $K^{\pm}$, and $p$+$\bar{p}$
  as a function of $p_{T}$ in 10-40\% centrality. This plot has been updated since the presentation\protect\footnotemark[2].}
\label{fig:pidv1}
\end{center}
\end{minipage}
\end{figure}

Figure~\ref{fig:vn} shows $v_{2}$, $v_{3}$, and $v_{4}$ of positive charged particles as a function
of $p_{T}$ using the event plane method and scalar product method.
Both methods are in a good agreement. Calculations from an
event-by-event viscous hydrodynamic model~\cite{bozek} are compared to
the data of $v_{2}$ and $v_{3}$. The model results using $\eta/s$=0.08
and $\eta/s$=0.16 qualitatively agree with the data in 0-5\% and
20-30\% centrality bins. The centrality dependence of $v_{n}$ are
similar to the results in Au+Au collisions~\cite{v2star,v3star,vnphenix}.
There was no significant difference between positive and negative charged particles for higher-order flows.

\begin{figure}[thb]
\begin{center}
\includegraphics[width=0.78\textwidth,angle=0,trim=0 0 0 0]{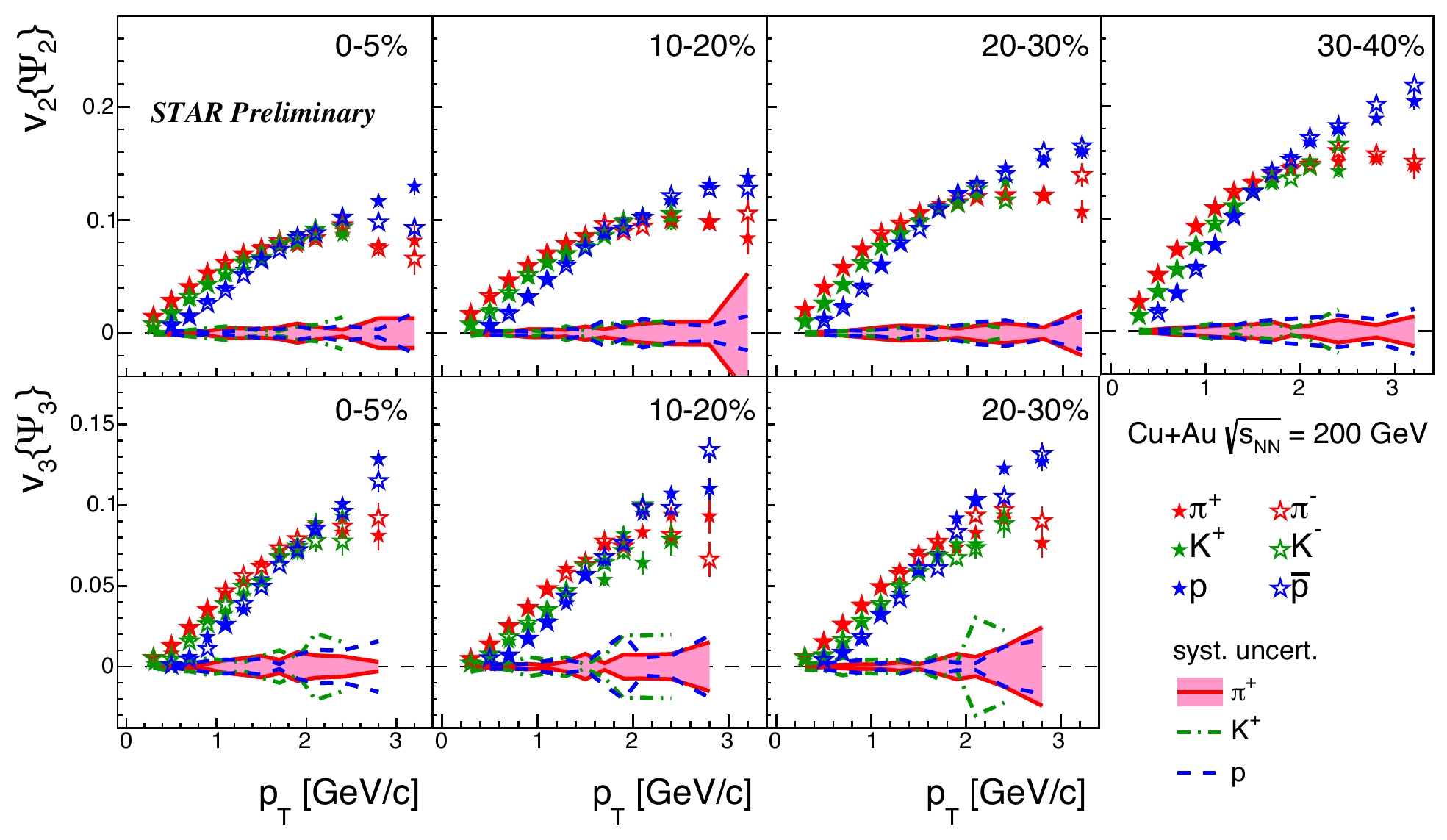}
\end{center}
\label{fig:pidvn}
\caption{$v_{2}$ and $v_{3}$ of $\pi$, $K$, and $p$($\bar{p}$) as a function of $p_{T}$
  for different centrality bins. This plot has been updated since the presentation\protect\footnotemark[2].}
\end{figure}

Identified particle $v_{n}$ are also measured combining the TPC dE/dx
and the time-of-flight information from Time-Of-Flight detector.  The
$v_{1}$ of charge-combined $\pi^{\pm}$, $K^{\pm}$, and $p+\bar{p}$ are
presented in Fig.~\ref{fig:pidv1} and the $v_{2}$ and
$v_{3}$ of $\pi^{+}(\pi^{-})$, $K^{+}(K^{-})$, and $p(\bar{p})$ for
different centrality bins are presented in 
Fig.~\ref{fig:pidvn}.  The same trends observed in symmetric
collisions, such as the mass ordering at low $p_{T}$ ($<$2 GeV/$c$) and the
baryon-meson splitting at intermediate $p_{T}$, are observed.

\section{Conclusions}
Charge-dependent anisotropic flow in Cu+Au collisions at
$\sqrt{s_{_{NN}}}$= 200 GeV has been  measured with the STAR detector.
Charge difference of $v_{1}$ is clearly observed, which is consistent
with the effect of the initial electric field. The magnitude of
$\Delta v_{1}$ is much smaller than the PHSD model predictions,
likely indicating that only a small fraction of all final state quarks
are created at the time when the electric field is strong. 
These results could shed light on the time
evolution of quark production in heavy-ion collisions. Higher-order
flow $v_{n}$ have been also presented; they exhibit similar trends observed in
symmetric collisions.
%




\bibliographystyle{elsarticle-num}
\bibliography{ref_qm15proc}

\begin{thebibliography}{10}
\expandafter\ifx\csname url\endcsname\relax
  \def\url#1{\texttt{#1}}\fi
\expandafter\ifx\csname urlprefix\endcsname\relax\def\urlprefix{URL }\fi
\expandafter\ifx\csname href\endcsname\relax
  \def\href#1#2{#2} \def\path#1{#1}\fi

\bibitem{hirono}
Y.~Hirono, M.~Hongo, T.~Hirano, Phys. Rev. C90 (2014) 021903.

\bibitem{voronyuk}
V.~Voronyuk, V.~D. Toneev, S.~A. Sergei, W.~Cassing, Phys. Rev. C90 (2014)
  064903.

\bibitem{2wave1}
S.~Pratt, PoS CPOD2013 (2013) 023.

\bibitem{2wave2}
B.~S.~A, P.~Danielewicz, S.~Pratt, Phys. Rev. Lett. 85 (2000) 2689.

\bibitem{TwoSub}
A.~M. Poskanzer, S.~A. Voloshin, Phys. Rev. C58 (1998) 1671.

\bibitem{SP}
{C. Adler {\it et al.} (STAR Collaboration)}, Phys. Rev. C66 (2002) 034904.

\bibitem{bozek}
P.~Bo$\dot{z}$ek, Phys. Lett. B717 (2012) 287--290.

\bibitem{v2star}
{J. Adams {\it et al.} (STAR Collaboration)}, Phys. Rev. C72 (2005) 014904.

\bibitem{v3star}
{J. Adams {\it et al.} (STAR Collaboration)}, Phys. Rev. C88 (2013) 014904.

\bibitem{vnphenix}
{A. Adare {\it et al.} (PHENIX Collaboration)}, Phys. Rev. Lett. 107 (2011)
  252301.

\end{thebibliography}







\end{document}